\documentstyle[12pt]{ws-art}
\begin{document}
\def\beq{\begin{equation}}
\def\eq{\end{equation}}
\def\etal{{\it et al.}}
\def\piC{$\pi^+$C$^{12}$}
\def\piPb{$\pi^+$Pb$^{208}$}
\def\sa{$S/A$ }
\def\pt{$p_T$ }
\def\et{$E_T$ }
\title{Subthreshold $K^+$ Production on Nuclei by $\pi^+$ Mesons}
\author{S.V. Efremov\\
{\it Bonner Nuclear Laboratory, Rice University, P.O. Box 1892,} \\
{\it Houston, TX 77251-1892, USA}
\and E.Ya. Paryev\\
{\it Institute for Nuclear Research, Russian Academy of Sciences,}\\
{\it Moscow 117312, Russia}}
\maketitle

\begin{abstract}
	The inclusive $K^+$ mesons production in $\pi^+$--nucleus reactions
	in the subthreshold energy regime is analyzed with respect to
	the one--step ($\pi^+n\to K^+\Lambda$) and the two--step
	($\pi^+n\to \eta p_1,~ \eta p_2\to K^+\Lambda$)
	incoherent production processes on the basis of an appropriate
	folding model, which allows one to take into account the various
 	forms of an internal nucleon momentum distribution as well as on--
	and off--shell propagation of the struck target nucleon. Contrary
        to proton--nucleus reactions primary reaction channel is found
        to be significant practically at all considered energies.
	Detailed predictions for the $K^+$ total and invariant
	differential cross sections from \piC--
	and \piPb--collisions at subthreshold energies
	are provided.
\end{abstract}
\newpage

\section*{Introduction}
	The extensive investigations of the production of $K^+$ mesons in
	proton--nucleus [1--7] and nucleus-nucleus [8--21] reactions at
	incident energies lower than the free nucleon-nucleon threshold
	have been carried out in the past years. Because of the high
	$K^+$ production threshold (1.58 $GeV$)
        in the nucleon-nucleon collision and the
	rather weak $K^+$ rescattering in the surrounding medium compared to
	the pions, etas, antiprotons and antikaons, from these studies one
        hopes to extract
 	some additional  information about the properties of nuclear matter,
	reaction dynamics, in--medium properties of hadrons at both
 	normal and high nuclear densities.
	However because of the complexity of collision  dynamics and
	uncertainties in elementary kaon
	production cross sections close to the production thresholds [17, 22],
 	in spite of large efforts,
	subthreshold kaon production is still far from being
	fully understood. To better understand the phenomenon
	of the subthreshold kaon production in pA-- and AA--interactions
	it is necessary to undertake experimental and theoretical
	investigation of the subthreshold kaon production in
 	$\pi A$--collisions, because in such
	collisions one may hope to get a clearer insight into the nuclear
	structure and the production mechanism [1, 23]. It should be noted
	that at present there is no experimental data on the kaon and antikaon
	subthreshold production in pion--induced reactions. Theoretical
	study of the subthreshold $K^-$ production in $\pi A$--collisions
	in the framework of the first--collision model has been performed
	elsewhere [23]. The aim of the present work is to explore the
	influence of an internal nucleon momentum distribution on the
	description of $K^+$ subthreshold production
	in $\pi^+A$--interactions
	as well as to evaluate the contributions from primary and
	secondary channels to the $K^+$ production process. In this study
	we present the predictions for total and differential $K^+$ production
	cross sections from \piC--  and \piPb--collisions
	in the subthreshold regime, obtained using the appropriate folding
	model for the primary and secondary production processes [23--25].

\section{The Model and Inputs}

\section*{1.1 Direct $K^+$ Production Process}

	Apart from participation in the elastic scattering an incident pion
	can produce a $K^+$ directly in the first inelastic
	$\pi N$--collision due to the nucleon Fermi motion.
	Since we are interested in the far subthreshold region,
	we have taken into account
	the following elementary process which requires the least amount of
	energy and, hence, has the lowest
	free production threshold (0.76 $GeV$)
	\footnote
{Calculations show that the elementary processes with a kaon
and a $\Sigma$--particle in the final states, opening up at slightly
higher energy (0.89 $GeV$), contribute to the total $K^+$ production
cross section on C$^{12}$ target nucleus with internal shell--model momentum
distribution at 700 $MeV$ initial kinetic energy the values of the order
of 8\% and 1\% when the target nucleon is on-- and off--shell,
respectively. The contributions from these processes and from
the primary process (1) to the kaon laboratory momentum spectrum
are comparable only at low kaon momenta. But we are
interested in the high--momentum part of the kaon spectrum
because of its strong sensitivity to the choice of the nucleon
momentum distribution}
	:

\beq
\pi^++n \to K^++\Lambda.
\eq
	Because the kaon rescatterings affect the spectrum mainly at large
	angles [12, 13, 26, 27] and since we have an intention to
	calculate the kaon spectrum at zero laboratory angle,
	where the cross section is
	the largest, we will neglect the kaon rescatterings
	in the present study
	\footnote
{This is allowed in calculating the total kaon yield as
kaons cannot be absorbed in nuclear medium due to the
strangeness conservation.}
	.
	Then we can represent the invariant inclusive cross section of $K^+$
	production on nuclei by the initial $\pi^+$ meson with momentum
	${\bf p}_0$ as follows [23--25]:

\beq
E_{K^+}\frac{d\sigma_{\pi^+A\to K^+X}^{(prim)}({\bf p}_0)}
{d{\bf p}_{K^+}}=
I_V[A,\sigma_{\pi^+N}^{tot}(p_0)]
\left<E_{K^+}\frac{d\sigma_{\pi^+n\to K^+\Lambda}({\bf p}_0,{\bf p}_{K^+})}
{d{\bf p}_{K^+}}\right>,
\eq
where
\beq
I_V[A,\sigma_{\pi^+N}^{tot}(p_0)]=N\int\rho({\bf r})d{\bf r}
\exp{[-\mu(p_0)\int\limits_{-\infty}^{0}\rho ({\bf r}+x{\bf \Omega}_0)dx]},
\eq
\beq
\mu(p_0)=\sigma_{\pi^+p}^{tot}(p_0)Z+\sigma_{\pi^+n}^{tot}(p_0)N;
\eq
\beq
\left<E_{K^+}\frac{d\sigma_{\pi^+n\to K^+\Lambda}({\bf p}_0,{\bf p}_{K^+})}
{d{\bf p}_{K^+}}\right>=
\int
n({\bf p}_t)d{\bf p}_t
\left[E_{K^+}
\frac
{d\sigma_{\pi^+n\to K^+\Lambda}(\sqrt{s}, {\bf p}_{K^+})}
{d{\bf p}_{K^+}}\right].
\eq
Here,
$E_{K^+}d\sigma_{\pi^+n\to K^+\Lambda}(\sqrt{s},{\bf p}_{K^+})/d{\bf p}_{K^+}$
is the free invariant inclusive cross section for the $K^+$
production in reaction (1);
$\rho({\bf r})$ and $n({\bf p}_t)$
are the density and ground--state momentum distribution
of intranuclear nucleons normalized to unit; ${\bf p}_t$
is the internal momentum of the struck target nucleon just before
the collision; $\sigma_{\pi^+N}^{tot}(p_0)$ is the total cross section of
free $\pi^+N$--interaction; Z and N are the numbers of protons and neutrons
in the target nucleus (A=N+Z); ${\bf \Omega}_0={\bf p}_0/p_0$;
${\bf p}_{K^+}$ and $E_{K^+}$ are the momentum and total energy of
a $K^+$ meson, respectively; $s$ is the $\pi^{+}n$ center--of--mass energy
squared.
The expression for $s$ is:
\beq
  s=(E_0+E_t)^2-({\bf p}_0+{\bf p}_t)^2,
\eq
  where $E_0$ and $E_t$ are the projectile's total energy, given by $E_0=
  \sqrt{p_{0}^{2}+m_{\pi}^{2}}$, and the struck target nucleon total energy,
  respectively. In our calculations we will use two formulas for $E_t$. In the
  first case we take into account the recoil of the residual nucleus. Then
  the energy that the struck target nucleon brings into the collision
  is equal to [3,28]:
\beq
   E_t=M_A-\sqrt{(-{\bf p}_t)^2+M_{A-1}^{2}},
\eq
 where $M_A$ and $M_{A-1}$ are the masses of the initial target nucleus
 and the recoiling residual nucleus, respectively. It is easily seen that
 in this case the struck target nucleon is off--shell and for a large target
 nucleus $E_t$ is approximately equal to the rest mass of nucleon $m_N$. In
 the opposite case we assume that the struck target nucleon
 is on--shell and
 $E_t$ is given simply by:
\beq
  E_t=\sqrt{p_{t}^{2}+m_{N}^{2}}.
\eq
Taking into account the two--body kinematics of the elementary process (1),
we can readily get the following expression for the Lorentz invariant inclusive
cross section for this process:
$$
E_{K^+}\frac{d\sigma_{\pi^+n\to K^+\Lambda}
(\sqrt{s},{\bf p}_{K^+})}
{d{\bf p}_{K^+}}=
\frac{\pi}
{I_2(s,m_{\Lambda},m_K)}
\frac{d\sigma_{\pi^+n\to K^+\Lambda}(s)}
{d\stackrel{*}{\bf \Omega}}\times
$$
\beq
\times
\frac{1}
{(\omega_1+E_t)}
\delta\left[\omega_1+E_t-\sqrt{m_{\Lambda}^2+({\bf Q}_1+{\bf p}_t)^2}
\right],
\eq
\beq
I_2(s,m_{\Lambda},m_K)=\frac{\pi}{2}
\frac{\lambda (s,m_{\Lambda}^2,m_K^2)}
{s},
\eq
\beq
\lambda(x,y,z)=\sqrt{[x-(\sqrt{y}+\sqrt{z})^2]
[x-(\sqrt{y}-\sqrt{z})^2]},
\eq
\beq
\omega_1=E_0-E_{K^+},~E_{K^+}=\sqrt{p_{K^+}^2+m_K^2},
{}~{\bf Q}_1={\bf p}_0-{\bf p}_{K^+}.
\eq
Here, $d\sigma_{\pi^+n\to K^+\Lambda}(s)/d\stackrel{*}{\bf \Omega}$
is the $K^+$ differential cross section in the $\pi^+n$ center--of--mass
system;
$m_\Lambda$, $m_K$ are the rest masses of a $\Lambda$  hyperon and a kaon.
According to Cugnon et al. [29], we choose the $K^+$ angular distribution
in the form, involving only Legendre polynomials up to the first order:
\beq
\frac{d\sigma_{\pi^+n\to K^+\Lambda}(s)}
{d\stackrel{*}{\bf{\Omega}}}=
[1+A_1(\sqrt{s})\cos{\stackrel{*}\vartheta_{K^+}}]
\frac{\sigma_{\pi^+n\to K^+\Lambda}(\sqrt{s})}
{4\pi}.
\eq
The parameter $A_1$ and the total cross section
$\sigma_{\pi^+n\to K^+\Lambda}$ of reaction (1)
can be parametrized by [29]:

\beq
A_1(\sqrt{s}) = \left\{
\begin{array}{ll}
	5.26\left(\frac{\sqrt{s}-\sqrt{s_0}}{{\rm GeV}} \right)
	& \mbox{for $\sqrt{s_0}<\sqrt{s}\le 1.8~{\rm GeV}$} \\
      1 & \mbox{for $\sqrt{s}>1.8~{\rm GeV}$},
\end{array}
\right.
\eq
\beq
\sigma_{\pi^+n\to K^+\Lambda}(\sqrt{s})=\left\{
\begin{array}{ll}
	10.0\left(\frac{\sqrt{s}-\sqrt{s_0}}{{\rm GeV}} \right)~[{\rm mb}]
	&\mbox{for $\sqrt{s_0}<\sqrt{s}\le 1.7~{\rm GeV}$} \\
	&\\
        0.09\left(\frac{{\rm GeV}}{\sqrt{s}-1.6 {\rm GeV}}\right)~[{\rm mb}]
	&\mbox{for $\sqrt{s}>1.7~{\rm GeV}$},
\end{array}
\right.
\eq
where $\sqrt{s_0}=m_K+m_{\Lambda}=1.61$ GeV is the threshold energy.

The internal nucleon momentum distribution, $n({\bf p}_t)$, is a crucial
point in the evaluation of the subthreshold production of any particles
on a nuclear target. Therefore, we calculated the cross sections
for the $K^+$ production in \piC-- and \piPb--
collisions using various types of
$n({\bf p}_t)$. For $K^+$ production calculations in the case of
C$^{12}$ target nucleus reported here the
following forms of $n({\bf p}_t)$ have been used.

The standard shell--model momentum distribution [23, 24]:
\beq
  n({\bf p}_t)=\frac{(b_0/\pi)^{3/2}}{A/4}\left\{1+\left[\frac{A-4}{6}\right]
b_0p_{t}^{2}\right\}\exp{(-b_0p_{t}^{2})},
\eq
where $b_0=68.5 ({\rm GeV/c})^{-2}$.

The momentum distribution in which the part corresponding to the $1p_{3/2}$
shell and having the exponential fall--off at high momentum $p_t$, was
inferred by Million [30] from the $(e,e'p)$ and $(\gamma, p)$ experiments [31]:
\beq
n({\bf p}_t)=\frac{1}{A/4}\left\{n_{1/2}({\bf p}_t)+\left[\frac{A-4}{4}
\right]n_{3/2}({\bf p}_t)\right\},
\eq
where
\begin{eqnarray}
n_{1/2}({\bf p}_t)&=&(\pi \nu)^{-3/2}\exp{(-p_t^2/\nu)},\\
n_{3/2}({\bf p}_t)&=&C_1n_{HO}({\bf p}_t)+C_2n_{exp}({\bf p}_t), \\
n_{HO}({\bf p}_t)&=&\frac{2}{3}(\pi \nu)^{-3/2}(p_t^2/\nu)\exp{(-p_t^2/\nu)},
	\nonumber \\
n_{exp}({\bf p}_t)&=&\left[\frac{1}{24{\pi}(p_t^0)^3}\right](p_t/p_t^0)
\exp{(-p_t/p_t^0)}
\end{eqnarray}
and $\sqrt{\nu}=127.0$ $MeV/c$, $p_t^0=55.0$ $MeV/c$, $C_1=0.997$, $C_2=0.003$.

The double Gaussian distribution with a large high--momentum tail extracted by
Geaga et al. [32] from high--energy proton backward scattering:
\beq
  n({\bf p}_t)=\frac{1}{(2\pi)^{3/2}(1+\alpha)}\left[\frac{1}{\sigma_{1}^{3}}
\exp{(-p_{t}^{2}/2\sigma_{1}^{2})}+
\frac{\alpha}{\sigma_{2}^{3}}\exp{(-p_{t}^{2}/2\sigma_{2}^{2})}\right],
\eq
 where $\sigma_1=0.119$
$GeV/c$, $\sigma_2=0.230$ $GeV/c$. The parameter $\alpha$ which defines the
 high--momentum part in $n({\bf p}_t)$ is 0.06 for C$^{12}$ and is proportional
 to $A^{1/3}$ for other target nuclei. It is worth noting that the
fractions of nucleons with intranuclear momenta greater than
the Fermi momentum $p_F=250$ $MeV/c$ are of the order of 10\%, 13\% and 25\%,
respectively for the distributions (16), (17) and (21). We show in Fig.1 the
$n({\bf p}_t)$ results for C$^{12}$. One can see that the momentum
distribution (21) differs significantly from distributions (16) and (17) at
$p_t\geq 400$ $MeV/c$.

For $K^+$ production calculations in the case of
\piPb--collisions besides the function (21) we used the following three types
of $n({\bf p}_t)$.

The completely degenerate Fermi gas momentum distribution:
\beq
  n({\bf p}_t)=\frac{1}{\frac{4}{3}{\pi}p_{F}^{3}}\theta{(p_F-p_t)},\,\,
 \theta{(x)}=\frac{x+|x|}{2|x|}.
\eq

Six Gaussian fit to the momentum distribution obtained from density-dependent
Hartree-Fock calculations with the Skyrme Hamiltonian [33]:

\beq
n({\bf p}_t)=
\sum_{i=1}^{6}n_i\left(\frac{b_i}
{\pi}\right)^{3/2}\exp{(-b_ip_t^2)},
\eq
where $(\sum_{i=1}^{6}n_i=1)$
$n=\{2.2745,-13.818,36.808,-40.447,19.868,-3.6855\}$,
$b_i=32.716\cdot 1.2804^i$ (GeV/c)$^{-2}$.

Moniz's parametrization based on calculations of nucleon--nucleon correlations
in the nuclear matter [34]:

\beq
n({\bf p}_t) = \left\{
\begin{array}{lll}
	( \frac{3}{4\pi p_F^3} )
 	\left(1-6\left(\frac{p_Fa_1}{\pi}\right)^2\right)
	&\mbox{for}&\mbox{$0<p_t<p_F$}   \\
	( \frac{3}{4\pi p_F^3} )
	\left(2\left(\frac{p_Fa_1}{\pi}\right)^2\left(\frac{p_F}{p_t}\right)^4
	 /\left(1-\frac{p_Fa_1}{8}\right)\right)
	&\mbox{for}&\mbox{$p_F<p_t<4~{\rm GeV/c}$} \\
       0&\mbox{for}&\mbox{$p_t>4 ~{\rm GeV/c}$},
\end{array}
\right.
\eq
with $a_1=2$(GeV/c)$^{-1}$. Use of the assumptions for $n({\bf p}_t)$
presented above enabled us to investigate the sensitivity of the predictions
for $K^+$ cross sections from $\pi^+A$--collisions to the high--momentum tail
of $n({\bf p}_t)$ at different incident energies.

	Consider now the integral (3) which represents the effective number
of neutrons for the $\pi^+n \to K^+\Lambda$ reaction on nuclei. A
simpler expression can be given for $I_V[A,\sigma_{\pi^+N}^{tot}(p_0)]$, if
$\rho ({\bf r})$ is a spherical function in the coordinate space [22]:
\beq
I_V[A,\sigma_{\pi^+N}^{tot}(p_0)]=\frac{\pi N}{\mu (p_0)}\int \limits_{0}^
{\infty}db_{\perp}^2 \left\{1-\exp
{\left[-\mu (p_0)\int \limits_{-\infty}^{\infty}
\rho \left(\sqrt{b_{\perp}^2+t^2}\right)dt\right]}\right\}.
\eq
In particular, for the Gaussian nuclear density
$(\rho({\bf r})=(b/\pi)^{3/2}\exp{(-br^2)}$, $b=0.248$ fm$^{-2}$ for
C$^{12}$) we get:
\beq
I_V[A,\sigma_{\pi^+N}^{tot}(p_0)]=\frac{N}{x_G}\int \limits_{0}^{1}
\frac{dt}{t}(1-e^{-x_Gt}),~~ x_G=\mu (p_0)b/\pi.
\eq
Whereas for a nucleus with a uniform density of nucleons of radius
$R$ the exact analytic expression for $I_V[A,\sigma_{\pi^+N}^{tot}(p_0)]$
can be readily obtained
\footnote{Numerical calculations, carried out in accordance with the formulas
(25)--(27), show that the difference between the results obtained
for a nucleus with a diffuse boundary and those obtained for
a nucleus with a sharp boundary constitutes a value of the order
of ten percents.}
:
\beq
I_V[A,\sigma_{\pi^+N}^{tot}(p_0)]=\frac{3N}{a_2^3}
\left[\frac{a_2^2}{2}-1+(1+a_2)e^{-a_2}\right],~~
a_2=\frac{3\mu (p_0)}{2\pi R^2}.
\eq
In our calculations of the $K^+$ production cross sections on Pb$^{208}$
and C$^{12}$ target nuclei we have used for $I_V[A,\sigma_{\pi^+N}^{tot}(p_0)]$
the equations (25) and (26), respectively. For the nuclear density
$\rho({\bf r})$ in the case of Pb$^{208}$ nucleus we have assumed
a Woods--Saxon form
$\rho({\bf r})=\rho_0/\left\{1+\exp{[(r-R_{1/2})/a]}\right\}$ with
$R_{1/2}=1.07A^{1/3}$ fm, $a=0.545$ fm [25].

Now let us perform an averaging of the $\pi^+n\to K^+\Lambda$ invariant
differential cross section (9) over the Fermi motion of the neutrons in
the nucleus using the properties of the Dirac $\delta$--function.
Choosing the spherical coordinates in ${\bf p}_t$ space with $z$ axis
parallel to ${\bf Q}_1$ and using the advantage of the spherical
symmetry of $n({\bf p}_t)$ we get the following formula for the
Fermi--averaged cross section (5) after the integration over the angle
between ${\bf Q}_1$ and ${\bf p}_t$ and a few other manipulations:
\beq
\left<E_{K^+}\frac{d\sigma_{\pi^+n\to K^+\Lambda}({\bf p}_0,{\bf p}_{K^+})}
{d{\bf p}_{K^+}}\right>=
\frac{\pi}{Q_1}\times
\eq
$$
\times\int \limits_{p_t^-}^{p_t^+}p_tdp_tn(p_t)
\int \limits_{0}^{2\pi}d\varphi
\frac{1}{I_2[s(x_0,\varphi),m_{\Lambda},m_K]}
\frac{d\sigma_{\pi^+n\to K^+\Lambda}[s(x_0,\varphi)]}{d\stackrel{*}\Omega},
$$
where
\beq
p_t^-=\left|Q_1\beta+\omega_1\sqrt{\beta^2+4m_N^2(Q_1^2-\omega_1^2)}\right|
/2(Q_1^2-\omega^2_1), ~~p_t^+=+\infty,
\eq
$$
Q_1=|{\bf Q}_1|,~~\beta = m_N^2-m_{\Lambda}^2-(Q_1^2-\omega_1^2);
$$
\beq
s(x,\varphi)=m_{\pi}^2+m_N^2+2E_0E_t-2p_0p_t
(\cos{\vartheta_{{\bf Q}_1}}\cdot x+
\sin{\vartheta_{{\bf Q}_1}}\cdot \sqrt{1-x^2}\cos{\varphi}),
\eq
\beq
x_0=(\beta+2\omega_1E_t)/2Q_1p_t,~
\cos{\vartheta_{{\bf Q}_1}}={\bf p_0}{\bf Q_1}/p_0Q_1,~
\sin{\vartheta_{{\bf Q}_1}}=\sqrt{1-\cos^2{\vartheta_{{\bf Q}_1}}}
\eq
in the cases when the struck target neutron is assumed to be on-shell
$(E_t=\sqrt{p_t^2+m_N^2})$ and
\footnote{The main contribution to the $K^+$ production in $\pi^+A$--collisions
at incident beam energies considered here comes from a region of the kaon
energies and momenta, where this condition is satisfied.}
$Q_1^2-\omega_1^2>0$, whereas
\beq
p_t^{\pm}=Q_1\pm\sqrt{(\omega_1+m_N)^2-m_{\Lambda}^2};
\eq
\beq
s(x,\varphi)=m_{\pi}^2+m_N^2+2m_NE_0-p_t^2-2p_0p_t
(\cos{\vartheta_{{\bf Q}_1}}\cdot x+
\sin{\vartheta_{{\bf Q}_1}}\cdot \sqrt{1-x^2}\cos{\varphi}),
\eq
\beq
x_0=\left[(\omega_1+m_N)^2-m_{\Lambda}^2-Q_1^2-p_t^2\right]/2Q_1p_t,
\eq
if the struck target neutron is assumed to be off--shell $(E_t=m_N)$.
It is worth noting that if the kaons escape in the beam direction then
the expression (28) for the Fermi--averaged invariant differential cross
section of the $\pi^+n\to K^+\Lambda$ process is reduced in view of
equations (30), (31) and (33) to an integral only over neutron momentum
$p_t$. Therefore, the differential cross section (2) for $K^+$ production
in $\pi^+A$--reactions in this case is more directly related to the
momentum distribution of the neutrons in the ground state than at other
outgoing angles. Figure 2 presents our calculations by (29) of the
lower limit of the relevant neutron's momenta at 600 and 700 $MeV$
primary--pion energies as a function of the momentum of a kaon produced
at an angle of 0$^0$ in the laboratory frame; and on Figure 3
the results of the similar calculations by (32) for the lower and
upper limits of these momenta are shown, when the struck target
neutron is assumed to be off--shell. It is clearly seen that in order
to produce kaons, for example, at 600 $MeV$ incident pion energy $\epsilon_0$
neutron momenta higher than about 200--250 $MeV/c$ are necessary.
Since in this momentum region the distribution function $n(p_t)$
falls rapidly with $p_t$ (see Fig.1), the integration of equation (28)
is exhausted within a relatively small range of $p_t$
values above $p_t^-$-- the smallest value of the neutron's momentum
$p_t$ which can satisfy the energy conservation condition in (9).
One can also see from Figure 2 that the lower is the pion energy,
the substantially larger internal momentum is necessary to produce a kaon
with a given high momentum in the on--shell struck target neutron case.
The allowed kinematical region becomes significantly smaller
with decreasing incident energy for the off--shell assumption about
the struck target neutron, as it follows from Figure 3.

	Finally, performing the integration of formula (2) (divided by
$E_{K^+}$) over the momentum ${\bf p}_{K^+}$ we easily get the following
expression for the total cross section for $K^+$ production
in $\pi^+A$--collisions
from the one--step elementary process (1)
\beq
\sigma_{\pi^+A\to K^+X}^{(prim)}({\bf p}_0)=
I_V[A,\sigma_{\pi^+N}^{tot}(p_0)]
\int n({\bf p}_t)d{\bf p}_t\sigma_{\pi^+n\to K^+\Lambda}(\sqrt{s}),
\eq
which is used hereafter.

	Let us focus now on the two--step $K^+$ production mechanism.

\section*{1.2 Two--Step $K^+$ Production Process}

	Kinematical considerations show that in the bombarding energy range
of our interest ($\epsilon_0\leq 0.76$ $GeV$) the following two--step
production
process may contribute to the $K^+$ production in $\pi^+A$--interactions
\footnote{We have neglected the two--step $K^+$ production processes
with heavier mesons in the intermediate states due to their larger
production thresholds in $\pi^+n$--collisions. For example, in the
case of $\omega$ meson the threshold energy is 0.96 $GeV$.}
. An incident pion can produce in the first inelastic collision with an
intranuclear neutron also an $\eta$ meson through the elementary reaction:
\beq
\pi^++n\to \eta +p.
\eq
We remind that the free threshold energy for this reaction is
0.56 $GeV$. Then the intermediate $\eta$ meson, which is assumed to be
on--shell, produces the kaon on a proton of the target nucleus via
the elementary subprocess with the lowest free production threshold
(0.20 $GeV$):
\beq
\eta + p\to K^++\Lambda,
\eq
provided that this subprocess is energetically possible. For instance,
the maximum kinetic energy of $\eta$ meson produced by a pion with the energy
$\epsilon_0=0.76$ $GeV$ on a target neutron at rest is about 0.33 $GeV$.
Therefore, for the beam energies considered here, there is a region of
eta's energy where the $K^+$ production process (37) occurs even if the
intranuclear proton is at rest.
Due to the Fermi motion of the protons the production
will be promoted by the target nucleus.
It is thus desirable to evaluate
the respective $K^+$ yield. In order to calculate the $K^+$ production
cross section for $\pi^+A$--reactions from the secondary eta induced
reaction channel (37) we fold the Fermi--averaged differential
cross section for the $\eta$ production in the reaction (36)
(denoted by $<d\sigma_{\pi^+n\to \eta p}({\bf p}_0,{\bf p}_{\eta})
/d{\bf p}_{\eta}>$) with the Fermi--averaged invariant differential
cross section for $K^+$ production in this channel
(denoted by $<E_{K^+}d\sigma_{\eta p\to K^+\Lambda}
({\bf p}_{\eta},{\bf p}_{K^+})/d{\bf p}_{K^+}>$)
and the effective number of np pairs per unit of square
(denoted by
$I_V[A,\sigma_{\pi^+N}^{tot}(p_0),\sigma_{\eta N}^{tot}(p_{\eta}),
\vartheta_{\eta}]$), i.e.:
\beq
E_{K^+}\frac
{d\sigma_{\pi^+A\to K^+X}^{(sec)}({\bf p}_0)}
{d{\bf p}_{K^+}}=
\int d{\bf p}_{\eta}
I_V[A,\sigma_{\pi^+N}^{tot}(p_0),
\sigma_{\eta N}^{tot}(p_{\eta}),\vartheta_{\eta}]\times
\eq
$$\times
\left<\frac{d\sigma_{\pi^+n\to \eta p}
({\bf p}_0,{\bf p}_{\eta})}{d{\bf p}_{\eta}}\right>
\left<E_{K^+}\frac{d\sigma_{\eta p\to K^+\Lambda}
({\bf p}_{\eta},{\bf p}_{K^+})}{d{\bf p}_{K^+}}\right>,
$$
where
\beq
I_V[A,\sigma_{\pi^+N}^{tot}(p_0),
\sigma_{\eta N}^{tot}(p_{\eta}),\vartheta_{\eta}]=
NZ\int \int d{\bf r}d{\bf r}_1\theta (x_{\|})
\delta^2({\bf x}_{\bot})\rho({\bf r})\rho({\bf r}_1)\times
\eq
$$\times
\exp{[-\mu(p_0)\int \limits_{-\infty}^{0}
\rho({\bf r}_1+x'{\bf \Omega}_0)dx'
-\mu(p_{\eta})\int \limits_{0}^{x_{\|}}
\rho({\bf r}_1+x'{\bf \Omega}_{\eta})dx']},
$$
\beq
{\bf r}-{\bf r}_1=x_{\|}{\bf \Omega}_{\eta}+{\bf x}_{\bot},
{\bf \Omega}_{\eta}={\bf p}_{\eta}/p_{\eta},
\mu(p_{\eta})=A\sigma_{\eta N}^{tot}(p_{\eta});
\eq
\beq
\left<\frac{d\sigma_{\pi^+n\to \eta p}({\bf p}_0,{\bf p}_{\eta})}
{d{\bf p}_{\eta}}\right>=
\int n({\bf p}_t)d{\bf p}_t\left[\frac
{d\sigma_{\pi^+n\to \eta p}(\sqrt{s},
{\bf p}_{\eta})}{d{\bf p}_{\eta}}\right],
\eq
\beq
\left<E_{K^+}\frac{d\sigma_{\eta p\to K^+ \Lambda}
({\bf p}_{\eta},{\bf p}_{K^+})}{d{\bf p}_{K^+}}\right>=
\int n({\bf p}_t)d{\bf p}_t
\left[E_{K^+}\frac{d\sigma_{\eta p\to K^+\Lambda}(\sqrt{s_1},{\bf p}_{K^+})}
{d{\bf p}_{K^+}}
\right]
\eq
and according to (9)
\beq
E_{\eta}\frac{d\sigma_{\pi^+n\to \eta p}(\sqrt{s},{\bf p}_{\eta})}
{d{\bf p}_{\eta}}=
\frac{\pi}{I_2(s,m_{\eta},m_N)}
\frac{d\sigma_{\pi^+n\to \eta p}(s)}{d\stackrel{*}\Omega}\times
\eq
$$\times
\frac{1}{(\omega_{\eta}+E_t)}
\delta\left[\omega_{\eta}+E_t-\sqrt{m_N^2+({\bf Q}_{\eta}+{\bf p}_t)^2}
\right],
$$
$$
\omega_{\eta}=E_0-E_{\eta},~~ {\bf Q}_{\eta}={\bf p}_0-{\bf p}_{\eta};
$$
\beq
E_{K^+}\frac
{d\sigma_{\eta p\to K^+\Lambda}(\sqrt{s_1},{\bf p}_{K^+})}
{d{\bf p}_{K^+}}=
\frac{\pi}{I_2(s_1,m_{\Lambda},m_{K})}
\frac{d\sigma_{\eta p\to K^+\Lambda}(s_1)}
{d\stackrel{*}\Omega}\times
\eq
$$\times
\frac{1}{(\omega_2+E_t)}
\delta \left[\omega_2+E_t-\sqrt{m_{\Lambda}^2+({\bf Q}_2+{\bf p}_t)^2}
\right],
$$
$$
\omega_2=E_{\eta}-E_{K^+},~~
{\bf Q}_2={\bf p}_{\eta}-{\bf p}_{K^+},~~
s_1=(E_{\eta}+E_t)^2-({\bf p}_{\eta}+{\bf p}_t)^2.
$$
Here, $d\sigma_{\pi^+n \to \eta p}/d\stackrel{*}{\Omega}$ and
$d\sigma_{\eta p\to K^+\Lambda}/d\stackrel{*}{\Omega}$ are the free
differential cross sections of the subprocesses (36) and (37)
in the corresponding center--of--mass systems;
$\sigma_{\eta N}^{tot}(p_{\eta})$ is the total cross section of the
free $\eta N$--interaction;
${\bf p}_{\eta}$, $E_{\eta}$ and $m_{\eta}$ are the momentum,
total energy and the rest  mass of $\eta$ meson, respectively,
$\cos{\vartheta_{\eta}}={\bf \Omega}_0{\bf\Omega}_{\eta}$.
The quantities $\mu (p_0)$, $s$ and $\theta(x)$ are defined above by
the equations
(4), (6) and (22), respectively. In our calculations the angular distributions
$d\sigma_{\pi^+n\to \eta p}/d\stackrel{*}{\Omega}$ and
$d\sigma_{\eta p\to K^+\Lambda}/d\stackrel{*}{\Omega}$
are assumed to be isotropic. For the total cross section
$\sigma_{\pi^+n\to \eta p}$ of the reaction (36) we used the
parametrization suggested in [35], assuming the isospin invariance
of the strong interaction ($\sigma_{\pi^+n\to \eta p}=
\sigma_{\pi^-p\to \eta n}$):
\beq
\sigma_{\pi^+n\to \eta p}(\sqrt{s}) = \left\{
\begin{array}{ll}
	13.07\left(\frac{\sqrt{s}-\sqrt{\stackrel{\sim}s_0}}
	{{\rm GeV}} \right)^{0.5288}~[{\rm mb}]
	&\mbox{for $\sqrt{\stackrel{\sim}{s}_0}<\sqrt{s}\le 1.589$ $GeV$}\\
	0.1449\left(\frac{{\rm GeV}}{\sqrt{s}-\sqrt{\stackrel{\sim}s_0}}
	\right)^{1.452}~[{\rm mb}]
	&\mbox{for $\sqrt{s}>1.1589$ $GeV$},
\end{array}
\right.
\eq
where $\sqrt{\stackrel{\sim}s_0}=1.486$ $GeV$ is the threshold energy.
Because of the lack of knowledge
\footnote{We know only one publication [36] in which
the authors basing on the SU(3)--symmetry of the strong
interaction obtained the following relation:
$\sigma_{\eta p\to K^+\Lambda}=
9/2\sigma_{\bar{K}^0p\to \Sigma^0\pi^+}$.
However, the use of this relation in the near threshold
energy region leads to very large contribution from the elementary
process (37) to the total $K^+$ production cross section
in $\pi^+A$--reactions, which is doubtful from our point of
view.}
about the total cross section $\sigma_{\eta p\to K^+\Lambda}$
of the elementary process (37), to estimate this cross section
we choose in
this work the following natural way, which was used also in [37]
for the evaluation of lambda production cross section in the reaction
$\omega N\to K\Lambda$. The probability for producing a kaon
in the reaction under consideration is given by
the ratio of the $\sigma_{\eta p\to K^+\Lambda}$ to the $\eta p$--inelastic
cross section $\sigma_{\eta p}^{in}$. We assume that this ratio
is equal to that of the $\sigma_{\pi^+n\to K^+\Lambda}$, defined by (15),
to the $\pi^+n$--inelastic cross section
$\sigma_{\pi^+n}^{in}$ at the same invariant energy $\sqrt{s_1}$.
Taking into account that $\sigma_{\eta p}^{in}\approx
\sigma_{\pi^+n}^{in}\approx 20$ mb [25, 38] in the eta and pion
energy ranges of interest, we get:
\beq
\sigma_{\eta p\to K^+\Lambda}(\sqrt{s_1})=
\sigma_{\pi^+n\to K^+\Lambda}(\sqrt{s_1}),
\eq
which is used hereafter to calculate the $K^+$ yield in $\pi^+A$--collisions
from the secondary channel (37).

Let us now simplify the expression (38) for the invariant
differential cross section for $K^+$ production in $\pi^+A$--interactions
from the two--step process. Taking into account that the main
contribution to the $K^+$ production comes from fast etas moving in the beam
direction and that the
$\eta N$ total cross section $\sigma_{\eta N}^{tot}$ in the energy
region of interest is approximately constant with a magnitude of
$<\sigma_{\eta N}^{tot}>\approx 35$ mb [25,35], we have:
\beq
E_{K^+}\frac{d\sigma_{\pi^+A\to K^+X}^{(sec)}({\bf p}_0)}
{d{\bf p}_{K^+}}=
I_V[A,\sigma_{\pi^+N}^{tot},<\sigma_{\eta N}^{tot}>,0^0]\times
\eq
$$\times
\int \limits_{0}^{\sqrt{E_0^2-m_{\eta}^2}}p_{\eta}^2dp_{\eta}
\left<\frac{d\sigma_{\pi^+n\to \eta p}({\bf p}_0,p_{\eta},0^0)}
{d{\bf p}_{\eta}}\right>
\left<
E_{K^+}\frac{d\sigma_{\eta p\to K^+\Lambda}
(p_{\eta}{\bf \Omega}_0,{\bf {\bf p}_{K^+}})}
{d{\bf p}_{K^+}}
\right>,
$$
where $<d\sigma_{\pi^+n\to \eta p}({\bf p}_0,p_{\eta},0^0)/d{\bf p}_{\eta}>$
is the corresponding spectrum of etas at an angle of $\vartheta_{\eta}=0^0$.
Using the way analogous to that used in the derivation
of equation (28) as well as after some algebra we can obtain:
\beq
\left<
\frac{d\sigma_{\pi^+n\to \eta p}({\bf p}_0,p_{\eta},0^0)}
{d{\bf p}_{\eta}}\right>=
\frac{\pi}{2E_{\eta}Q_{\eta}}
\int \limits_{p_G^-}^{p_G^+}p_tdp_tn(p_t)
\frac{\sigma_{\pi^+n\to \eta p}[\sqrt{s(y_0)}]}
{I_2[s(y_0),m_{\eta},m_N]},
\eq
where
\beq
p_G^-=\left|\frac{1}{2}Q_{\eta}-\omega_{\eta}\sqrt{\frac{1}{4}+
\frac{m_N^2}{Q_{\eta}^2-\omega_{\eta}^2}}\right|,~~
p_G^+=+\infty,
\eq
$$
s(y)=m_{\pi}^2+m_N^2+2E_0E_t-2p_0p_ty,~~
y_0=(\omega_{\eta}^2-Q_{\eta}^2+2\omega_{\eta}E_t)/(2Q_{\eta}p_t)
$$
in the case when the struck target nucleon is assumed to be on--shell and
\beq
p_G^{\pm}=\left|Q_{\eta}\pm \sqrt{\omega_{\eta}^2+2m_N\omega_{\eta}}\right|,
\eq
$$
s(y)=m_{\pi}^2+m_N^2+2m_NE_0-p_t^2-2p_0p_ty,~~
y_0=(\omega_{\eta}^2+2m_N\omega_{\eta}-Q_{\eta}^2-p_t^2)/(2Q_{\eta}p_t)
$$
if the struck target nucleon is off--shell. The quantity
$<E_{K^+}\frac
{d\sigma_{\eta p\to K^+\Lambda(p_{\eta}{\bf \Omega}_0,{\bf p}_{K^+})}}
{d{\bf p}_K^+}>$ in (47) is determined by the formulas (28)--(34) in which
we have to make the following substitutions:
$$
d\sigma_{\pi^+n\to K^+\Lambda}/d\stackrel{*}\Omega \to
d\sigma_{\eta p\to K^+\Lambda}/d\stackrel{*}\Omega,
$$
$$
s \to  s_1,
$$
$$
{\bf Q}_1, \omega_1 \to {\bf Q}_2, \omega_2,
$$
$$
E_0, {\bf p}_0 \to E_{\eta},p_{\eta}{\bf \Omega}_0
$$
$$
m_{\pi}\to m_{\eta}
$$
and to take into account that for on--shell assumption about the struck
target proton if $Q_2^2-\omega_2^2<0$ (this case is realized here) then
the lower and upper limits in (28) are:
\beq
p_t^{\pm}=\left| Q_2\beta \mp\omega_2\sqrt{\beta^2+4m_N^2(Q_2^2-\omega_2^2)}
\right|/2(\omega_2^2-Q_2^2).
\eq

One can show that the expression for
$I_V[A,\sigma_{\pi^+N}^{tot}(p_0),<\sigma_{\eta N}^{tot}>,0^0]$
in the case  of a nucleus of a radius $R=1.3A^{1/3}$ fm with a sharp
boundary has the following simple form:
\beq
I_V[A,\sigma_{\pi^+N}^{tot}(p_0),<\sigma_{\eta N}^{tot}>,0^0]=
\frac{9NZ}{2\pi R^2(a_2-a_3)}\times
\eq
$$\times
\left \{
\frac{1}{a_2^3}[1-(1+a_2)e^{-a_2}]-\frac{1}{a_3^3}[1-(1+a_3)e^{-a_3}]-
\frac{1}{2a_2}+\frac{1}{2a_3}
\right \},
$$
where $a_3=3A<\sigma_{\eta N}^{tot}>/2\pi R^2$ and the quantity $a_2$
is determined by the formula (27).

	Besides of the differential cross section it is of further interest
to get the corresponding expression for the total cross section for $K^+$
production in $\pi^+A$--reactions from the two--step process. Integrating
the formula (47) (divided by $E_{K^+}$), we readily obtain
\beq
\sigma_{\pi^+A\to K^+X}^{(sec)}({\bf p}_0)=
I_V[A,\sigma_{\pi^+N}^{tot}(p_0),<\sigma_{\eta N}^{tot}>,0^0]\times
\eq
$$\times
\int \limits_{0}^{\sqrt{E_0^2-m_{\eta}^2}}p_{\eta}^2dp_{\eta}
<\frac{d\sigma_{\pi^+n\to \eta p}({\bf p}_0,p_{\eta},0^0)}{d{\bf p}_{\eta}}>
\int n({\bf p}_t)d{\bf p}_t\sigma_{\eta p\to K^+\Lambda}(\sqrt{s_1}).
$$

	Now let us discuss the results of our calculations in the
framework of the approach outlined above.

\section{Results and Discussion}

The expected total cross sections for $K^+$ production in $\pi^{+}+C^{12}$
--reactions from the primary $\pi^{+}n\rightarrow K^+\Lambda$ and secondary
$\eta p\rightarrow K^+\Lambda$ channels calculated according to (6)--(8),
(15)--(21), (26), (35), (45), (46), (48)--(50), (52), (53) are shown in
Fig.4 as functions of the laboratory energy $\epsilon_0$ of the pion.
The elementary cross sections $\sigma_{{\pi^{+}}N}^{tot}$ in the calculations
 were borrowed from [38]. It is seen that the $\eta$ induced production
channel becomes comparable to the $\pi^{+} n$ channel only at very low
energies ($\epsilon_0 < 600$ $MeV$) if we adopt the off--shell assumption
about the struck target nucleon as well as use the nucleon momentum
distributions (16), (17) without a large high--momentum tails. The cross
section
of the $K^+$ production from the primary $\pi^{+}n$ production channel in the
energy region of $600\le\epsilon_0\le 650$ $MeV$, where this channel dominates,
still strongly depends both on the choice of the nucleon momentum distribution
and on the bombarding energy $\epsilon_0$ (contrary to the secondary production
channel). The cross section calculated with the momentum distribution (21),
extracted from high--energy proton backward scattering, is larger by a factor
of 2--10 than that calculated with the shell--model momentum distribution (16)
in this region. The momentum distribution (17), deduced from the ($e,e'p$)
and ($\gamma,p$) experiments, gives practically the same results as the
shell--model momentum distribution (16). The values of the total production
cross section in the energy region under consideration lie in the range of
0.5--30 $\mu$b. Such rapid energy dependence of the ($\pi^+,K^+$) total cross
section in the energy region considered here is a characteristic signature of
the $\pi^{+}n\rightarrow K^+\Lambda$ one--step production mechanism. Therefore,
the measurement of the total $K^+$ production cross section at incident
energies between 600 and 650 $MeV$ seems to be quite
promising to study both the kaon
production mechanism and the high--momentum components within nucleus. However,
it is important to emphasize that in order to get clearer insight into the
relative role of the primary and secondary reaction channels, further
theoretical efforts are necessary for a better understanding about the $\eta$
induced elementary $K^+$ production process (37).

Figure 5 presents the results of similar calculations, using the momentum
distributions (21)--(24), for the total cross section for $K^+$ production
in $\pi^{+}+Pb^{208}$--reactions. It is seen that the contribution from the
two--step production process is comparable to that from the one--step
production process, as well as in the previous case, only at energies
$\epsilon_0 < 600$ $MeV$. In order to demonstrate the sensitivity of the kaon
yield from the two--step production process to $\eta$ absorption in the nuclear
medium, in Fig. 5 we show the results of calculations carried out under the
assumption that the total cross section of $\eta N$--interaction is replaced
by the total inelastic cross section of this interaction with a magnitude of
20 mb. We can observe an increase of the $K^+$ yield by a factor of 1.5 for
this target nucleus. It is also seen that the cross section for $K^+$
production in the far subthreshold region ($600\le\epsilon_0\le 650$ $MeV$) in
this case is larger by a factor of 10 than that calculated for
$\pi^{+}+C^{12}$--reaction. One can see further that for the one--step
production process (1) to occur at the bombarding energies
$\epsilon_0 < 600$ $MeV$, the internal momenta greater
than 250 $MeV/c$ are needed,
if the struck target neutron is assumed to be off-shell. This
finding is in agreement with that of Figure~3. It is interesting to note that
the nucleon momentum distributions (21) and (24) having the different forms of
a high--momentum tail give the close results for the $K^+$ total cross section
from the primary production channel. Therefore, in order to differentiate
between these distributions one needs, as it follows from the expression (28),
to obtain information about the differential cross sections of the $K^+$
mesons produced in the reactions under consideration.

In Figs.6 and 7 we show the results of our calculations by (2)--(26),
(28)--(34), (38)--(52) of the inclusive invariant cross sections for the
kaon production from the primary $\pi^{+}n$-- and secondary $\eta p$--reaction
channels at an angle of 0$^0$ in the interaction of pions with the energies
of 600 $MeV$ (lower lines) and 700 $MeV$ (upper lines) with $C^{12}$ and
$Pb^{208}$ nuclei. It is clearly seen that the secondary production channel
practically does not contribute to the spectrum of emitted kaons at incident
energies between 600 and 700 $MeV$. Also one can see that the high--momentum
part of the spectrum of kaons from the one--step production process (1) is
essentially determined by the nucleon momentum distribution, whereas its
low--momentum part is not sensitively affected by the choice of the internal
momentum distribution in the case of the on--shell assumption about the struck
target neutron. The difference between the calculations with allowance for
the high--momentum components in the target nucleus and without it becomes
larger at lower incident energies also in the case when the struck target
neutron is off--shell. Taking into account the considered above, we conclude
that the measurement of the total and differential $K^+$ production cross
sections at incident energies between 600 and 700 $MeV$ offers the possibility
to check the dominant role of the one--step production process in the
subthreshold $K^+$ production as well as to study the high--momentum components
within target nucleus.

\section{Summary}

In this paper we have calculated the total and differential cross sections
for $K^+$ production from $\pi^{+}$+C$^{12}$--
and $\pi^{+}$+Pb$^{208}$--reactions
in the subthreshold regime by considering incoherent primary pion--neutron and
secondary eta--proton production processes within the framework of an
appropriate folding model. This approach takes into account the various forms
of an internal nucleon momentum distribution as well as on-- and off--shell
propagation of the struck target nucleon. It was shown that, contrary to
proton--nucleus reactions, the one--step $K^+$ production mechanism clearly
dominates at all subthreshold energies if the struck target nucleon is assumed
to be on--shell, whereas for the off--shell assumption about the struck target
nucleon the one--step and the two--step reaction mechanisms are of equal
importance
only at very low energies ($\epsilon_0 < 600$ MeV) in the case of use of the
nucleon momentum distributions without a large high--momentum tails. We
predict, in particular, a rapid energy dependence of the total cross section
for $K^+$ production from the primary channel as well as its strong sensitivity
to the choice of the internal nucleon momentum distribution in the energy
domain
$\epsilon_0\approx 600 \div 650$ $MeV$, where this channel dominates. Such
behaviour of the ($\pi^{+},K^+$) total cross section in this energy domain is a
principal signature of the $\pi^{+}n\rightarrow K^+\Lambda$ production
mechanism. Therefore, from measurements of the inclusive $K^+$ production cross
sections at incident pion energies indicated above one may hopes to obtain a
clear information both on the production mechanism and on the high--momentum
components within target nucleus.
\centerline{ }
\centerline{\bf Acknowledgements}
\centerline{ }
We would like to thank V.Koptev for stimulating discussions on the initial
stage of this study. We are also grateful to K.Oganesyan for interest in the
work. One of us (E.Ya.P.) was supported in part by Grant No.N6K000 from the
International Science Foundation as well as by Grant No.N6K300 from the
International Science Foundation and Russian Government.

\newpage

\centerline{\bf Figure captions}
\centerline{ }
{\bf Fig.1.} Momentum distribution for $C^{12}$. The solid line is the
  harmonic oscillator momentum distribution (16). The dot--dashed line is the
  momentum distribution (17), extracted from the $(e,e'p)$
  ($p_t<$ 250 $MeV/c$) and $(\gamma,p)$ ($p_t > $250 $MeV/c$) experiments [31].
  The dashed line is the momentum distribution (21) inferred from
  high--energy proton backward scattering [32].\\

{\bf Fig.2.} Lower limit of the internal neutron momentum as a function of the
  momentum of a kaon produced in beam direction at 600 and 700 $MeV$ incident
  pion energies in the case when the target neutron is assumed to be
  on--shell.\\

{\bf Fig.3.} Lower and upper limits of the internal neutron momentum as a
  function of the momentum of a kaon produced in beam direction at 600 and
  700 $MeV$ incident pion energies in the case when the target neutron is
  assumed to be off--shell.\\

{\bf Fig.4.} The total cross sections for $K^+$ production in
  $\pi^++C^{12}$--reactions from primary $\pi^{+}n\rightarrow K^+\Lambda$
  and secondary $\eta p\rightarrow K^+\Lambda$ channels
  as functions of the laboratory energy of the pion. The cross sections from
  primary $\pi^{+}n$--collisions: the heavy and the light solid lines
  are calculations with the shell--model momentum distribution (16) with
  off-- and
  on--shell assumptions about the struck target neutron, respectively;
  the dot--dashed line is the calculation with the momentum distribution (17)
  and off--shell struck target neutron;
  the short-- and the long--dashed lines are calculations with distribution
(21)
  with large high--momentum tail with off-- and on--shell
  assumptions about the struck target neutron, respectively. The cross sections
  from secondary $\eta p$--collisions: the lower and upper dotted lines are
  calculations with the shell--model momentum distribution (16) with off--
  and on--shell assumptions about the struck target nucleons, respectively;
  the short-- and the long--dashed lines with two dots are calculations with
  distribution (21) with off-- and on--shell assumptions about the struck
  target nucleons, respectively. The arrows show the
  production thresholds on a free neutron and on off--shell neutron with the
  Fermi momentum of 250 $MeV/c$ and the absolute production threshold.\\

{\bf Fig.5.} The calculated total cross sections for $K^+$ production in
  $\pi^++Pb^{208}$--reactions from primary $\pi^{+}n$--
  and secondary $\eta p$-- collisions
  as functions of the laboratory energy of the pion. The cross sections from
  primary $\pi^{+}n$--collisions: the heavy and the light solid lines
  are calculations with the Fermi gas momentum distribution (22) with
  off-- and
  on--shell assumptions about the struck target neutron, respectively;
  the short--long--dashed line is the calculation with six Gaussian
  distribution (23) and off--shell struck target neutron; the short-- and the
  long--dashed lines with one dot are calculations with the Moniz's
  parametrization (24) with off-- and on--shell
  assumptions about the struck target neutron, respectively. The cross sections
  from secondary $\eta p$--collisions: the lower and upper dotted lines are
  calculations with the Fermi gas momentum distribution (22) with off--
  and on--shell assumptions about the struck target nucleons, respectively;
  the crosses denote the same as lower dotted line, but it is supposed that
  the total cross section of $\eta N$--interaction is replaced by its total
  inelastic cross section with a magnitude of 20 mb. The rest of the notation
  is the same as in Figure~4.\\

{\bf Fig.6.} The inclusive invariant cross sections for the production of $K^+$
  mesons in primary $\pi^{+}n$-- and secondary $\eta p$--collisions at an
  angle of 0$^0$ as functions of the kaon momentum in the interaction of pions
  with the energies of 600 and 700 $MeV$ with $C^{12}$ nuclei. The dot--dashed
  line denotes the same as in Figure~4, but it is supposed that the struck
  target neutron is on--shell. The rest of the notation is the same as in
  Figure~4.\\

{\bf Fig.7.} The inclusive invariant cross sections for the production of $K^+$
  mesons in primary $\pi^{+}n$-- and secondary $\eta p$--collisions at an
  angle of 0$^0$ as functions of the kaon momentum in the interaction of pions
  with the energies of 600 and 700 $MeV$ with $Pb^{208}$ nuclei. The
  short--short--long--dashed
  line denotes here the same as the short--long--dashed line, but it is
  supposed that the struck target neutron is on--shell. The rest of the
  notation is the same as in Figure~5.\\

\end{document}